\documentclass[fleqn,twoside,twocolumn,nofootinbib,showkeys]{revtex4} 
\usepackage[sec,nocpr]{ujp} 

\def\lsim{\
  \lower-1.2pt\vbox{\hbox{\rlap{$<$}\lower5pt\vbox{\hbox{$\sim$}}}}\ }
\def\gsim{\
  \lower-1.2pt\vbox{\hbox{\rlap{$>$}\lower5pt\vbox{\hbox{$\sim$}}}}\ }

\begin{document}
\title[Electric field and electric forces in a spontaneously polarized nonpolar isotropic dielectric]
{ELECTRIC FIELD AND ELECTRIC FORCES \\ IN A SPONTANEOUSLY POLARIZED \\ NONPOLAR ISOTROPIC DIELECTRIC}%
\author{Maksim D.~Tomchenko}
\affiliation{Bogolyubov Institute for Theoretical Physics, Nat. Acad. of Sci. of Ukraine}
\address{14b, Metrolohichna Str., Kyiv 03143, Ukraine}
\email{mtomchenko@bitp.kiev.ua}

\udk{538.956}  \razd{\secvii}

\autorcol{Maksim D.~Tomchenko}

\setcounter{page}{1}%

\begin{abstract}
Based on the microscopic Maxwell equations, we develop a method of
description of the electric field in a spontaneously polarized
isotropic nonpolar dielectric.  We  find the solution for the
electric field $\textbf{E}(\textbf{r})$ for several typical
examples. Moreover, we generalize Helmholtz's formula for the
electric force acting on a volume element of a dielectric with
regard for the contribution of the spontaneous polarization.
\end{abstract}

\maketitle

\section{Introduction}
As known, the polarization of dielectrics can be of two types: the
polarization induced by an external electric field and the
spontaneous one
\cite{maxwell,smythe,jackson,land8,tamm,born,strukov}. The
spontaneous polarization is characteristic of ferroelectrics,
pyroelectrics, and some of piezoelectrics. It can also arise in
``customary'' dielectrics. In particular, electric signals were
observed in He II under torsion oscillations \cite{ryb2005} and in
standing half-waves of the second \cite{ryb2004,chag2016,yayama2018}
and first \cite{chag2017} sounds. In those experiments, the electric
field can be associated only with the spontaneous polarization,
since $^{4}He$ atom possesses no intrinsic charge, dipole moment or
multipole moment.

Depending on the type of a dielectric, its atoms can be charged or
neutral (the latter can be polar or nonpolar). For the ionic
crystals, which can be piezoelectics, pyroelectrics, or
ferroelectrics under certain conditions, the theory of spontaneous
polarization is well developed
\cite{land8,born,strukov,tagantsev1987,ginzburg2001,moroz2009}.
However, no such theory is available for nonpolar dielectrics (to
our know\-ledge), though the properties of nonpolar dielectrics are
more simple. In what follows we try to remove this gap. We will
consider only isotropic dielectrics.

For the nonpolar dielectrics, one needs to represent the total
polarization $\textbf{P}(\textbf{r})$ as
$\textbf{P}_{i}(\textbf{r})+\textbf{P}_{s}(\textbf{r})$ and to
consider $\textbf{P}_{s}$ and $\textbf{P}_{i}$ self-consistently
(here, $\textbf{P}_{s}$ is the ``bare'' spontaneous polarization,
and $\textbf{P}_{i}$ is the induced polarization). This is seen from
the following. In a nonpolar isotropic dielectric, the spontaneous
polarization can arise due to sound, gravity, acceleration, and
elastic deformation, since a directed concentration gradient appears
in the medium in these cases. A nonpolar atom has no intrinsic
dipole moment. However, two nonpolar atoms polarize each other. As a
result, each atom acquires the dipole moment \cite{wb1,wb2,lt2011}
\begin{equation}
 \textbf{d} = -D_{7} |e| \frac{a_{B}^{8}}{r^7} \textbf{i}_{\textbf{r}},   \label{0-0}     \end{equation}
where $r$ is the distance between atoms, $\textbf{i}_{\textbf{r}}
=\textbf{r}/r$ is the unit vector directed to another atom, $a_B =
\frac{\hbar^2}{me^2}$ is the Bohr radius, and $D_{7}$ is an atomic
constant. In view of this, the gradient of the particle number
density $n(\textbf{r})$ causes the bulk polarization of the medium
\begin{equation}
  \textbf{P}_{s}(\textbf{r}) = \xi\nabla n(\textbf{r}).   \label{0-1}     \end{equation}
The rough estimate on the basis of formula (\ref{0-0}) gives
$\xi=(7/3)d_{0}\bar{r}_{0}(n(\textbf{r})/n_{0})^{a}$, where $a=2$,
$d_{0} =-D_{7} |e| \frac{a_{B}^{8}}{\bar{r}_{0}^7}$, and
$\bar{r}_{0}=n_{0}^{-1/3}$ is the mean interatomic distance in the
unperturbed system. The more accurate analysis with the averaging of
$\textbf{d}$ over different configurations of atoms leads to
$\xi\approx 7.5(7/3)d_{0}\bar{r}_{0}(n(\textbf{r})/n_{0})^{a}$,
where $a=1$ \cite{mt2010} (the number $7.5$ is true only for He II,
the sign of $\xi$ was obtained in \cite{mt2010} improperly). The
collection of dipoles creating the polarization
$\textbf{P}_{s}(\textbf{r})$ induces the field
$\textbf{E}(\textbf{r})$. This field additionally polarizes the
atoms and causes the induced polarization
$\textbf{P}_{i}(\textbf{r})$. The latter was not taken into account
in $\textbf{P}_{s}(\textbf{r})$ and also affects
$\textbf{E}(\textbf{r})$. Thus, the electron shell of each atom is
deformed by two different forces: first, due to the quantum mutual
polarization of atoms and the density gradient, second, due to the
field $\textbf{E}$ in the medium. It is clear that, at small
deformations, those two deformations should be independent of each
other. At large deformations we need to solve a quantum-mechanical
problem of two nonpolar atoms in the external field $\textbf{E}$.
Such problem has not been solved yet, as far as we know. In what
follows, we consider the deformations to be small. In this case, the
polarizations $\textbf{P}_{s}$ and $\textbf{P}_{i}$ should be
considered separately. We believe that this is true for all
mechanisms of spontaneous polarization of nonpolar (isotropic or
anisotropic) dielectrics.

Based on microscopic Maxwell  equations, we will develop a method of
determination  of the polarization
$\textbf{P}(\textbf{r})=\textbf{P}_{s}(\textbf{r})+\textbf{P}_{i}(\textbf{r})$
and the field $\textbf{E}(\textbf{r})$ for isotropic nonpolar
dielectrics with the self-consistent account for $\textbf{P}_{s}$
and $\textbf{P}_{i}$. We will also show that the consideration of
$\textbf{P}_{s}$ changes the force acting on an element of the
volume of a dielectric.

\section{Electric field}
In the electrodynamics of continua (see \cite{land8}, Chapters II,
IV, IX) the electromagnetic field of isotropic dielectrics is
described by the local (i.e., valid at each point of a medium)
relations
\begin{equation}
 \textbf{D}=\textbf{E}+4\pi \textbf{P}=\varepsilon\textbf{E},
       \label{1} \end{equation}
\begin{equation}
 \textbf{B}=\textbf{H}+4\pi \textbf{M} =\mu\textbf{H}
     \label{4} \end{equation}
and by the Maxwell equations in a medium:
\begin{equation}
 div \textbf{D}=4\pi \rho_{f},
       \label{5} \end{equation}
\begin{equation}
rot \textbf{E}=-\frac{1}{c}\frac{\partial \textbf{B}}{\partial t},
        \label{6} \end{equation}
\begin{equation}
 div \textbf{B}=0,
       \label{7} \end{equation}
\begin{equation}
rot \textbf{H}=\frac{1}{c}\frac{\partial \textbf{D}}{\partial
t}+\frac{4\pi \textbf{j}_{f}}{c}.
       \label{8} \end{equation}
Here, $\rho_{f}$ is the density of foreign charges, and
$\textbf{j}_{f}$ is the foreign current.

The spontaneous polarization of a dielectric arises due to the
internal ``nonelectric'' mechanism causing the appearance of a set
of dipoles in the medium; we will call them ``spontaneous'' dipoles.
We will see below that, in order to describe a spontaneously
pola\-rized nonpolar isotropic dielectric, one needs to make some
changes in Eqs. (\ref{1})--(\ref{8}). Let us deduce the necessary
equations from the microscopic Maxwell equations.

First, we find the formula for the polarization
$\textbf{P}(\textbf{r})$, since the formulae available in the
literature are not quite accurate sometimes.  The density of
polarization charges $\bar{\rho}(\textbf{r}),$ averaged over the
physically infinitesimal volume, satisfies the relation
\begin{equation}
\int\limits_{V_{+}}\bar{\rho}(\textbf{r})dV=0.
       \label{d1} \end{equation}
Therefore, $\bar{\rho}(\textbf{r})$ can be presented in the form
$\bar{\rho}(\textbf{r})=-div \textbf{P}(\textbf{r})$, where
$\textbf{P}=0$ outside the dielectric. Here, $V_{+}$ is the volume
confined inside the surface that co\-vers the whole dielectric and
oversteps its limits by an infinitesimal distance. In \cite{land8}
(Chapt. 2, $\S 6$) $\textbf{P}(\textbf{r})$ is determined with the
use of the following formulae:
\begin{eqnarray}
&&\int\limits_{V_{+}}\textbf{r}\bar{\rho}(\textbf{r})dV =
-\int\limits_{V_{+}}\textbf{r}
(\nabla\textbf{P}(\textbf{r}))dV=\nonumber \\&&=
-\oint\limits_{S}\textbf{r} (\textbf{P}(\textbf{r})d\textbf{S})+
\int\limits_{V_{+}}\textbf{P}(\textbf{r})dV=\int\limits_{V_{+}}\textbf{P}(\textbf{r})dV.
       \label{d2} \end{eqnarray}
However, these formulae do not allow us to find
$\textbf{P}(\textbf{r})$, since relation (\ref{d2}) does not yield
$\textbf{P}(\textbf{r})=\textbf{r}\bar{\rho}(\textbf{r})$. Indeed,
let the polarization be uniform:
$\textbf{P}(\textbf{r})=\textbf{P}$. Then
$\bar{\rho}(\textbf{r})=-div \textbf{P}=0$ inside the body. On the
surface $\textbf{P}(\textbf{r})$ decreases by jump to zero, and
$\bar{\rho}(\textbf{r})$ is singular. For such dielectric, the
equality $\textbf{P}(\textbf{r})=\textbf{r}\bar{\rho}(\textbf{r})$
is violated at all points of the body. In this case, the integral
$\int_{V_{+}}\textbf{r}\bar{\rho}(\textbf{r})dV$ is defined by the
surface part of $\bar{\rho}(\textbf{r})$. These properties are
natural, since a uniformly polarized body can be considered as two
bo\-dies that possess uniformly distributed $(+)$ and $(-)$ charges
and are shifted relative to each other by an infinitesimal distance.
The jump of $\bar{\rho}(\textbf{r})$ on the surface is usually large
for the nonuniform polarization as well. In this case, the equality
$\textbf{P}(\textbf{r})=\textbf{r}\bar{\rho}(\textbf{r})$  should
also be significantly violated.  We note that the de\-finition
$\textbf{P}(\textbf{r})=\textbf{r}\bar{\rho}(\textbf{r})$
\cite{land8} was criticized previously in work \cite{tagantsev1987}.

In order to obtain the formula for $\textbf{P}(\textbf{r})$ we note
that, for a polarized dielectric without foreign charges,
$\rho(\textbf{r})$ reads
\begin{equation}
 \rho(\textbf{r})=\sum\limits_{j=1}^{N}[q^{(j)}\delta(\textbf{r}-\textbf{r}_{j})-q^{(j)}
 \delta(\textbf{r}-\textbf{r}_{j}-\textbf{r}_{0}^{(j)})],
       \label{d3} \end{equation}
where $N$ is the number of atoms in the dielectric. We set
$q^{(j)}<0$ for all $j$. Then
\begin{equation}
 \int\limits_{V_{+}}\textbf{r}\rho(\textbf{r})dV=-\sum\limits_{j=1}^{N}
 \textbf{r}_{0}^{(j)}q^{(j)}=\sum\limits_{j=1}^{N}
 \textbf{d}_{j}=\int\limits_{V_{+}}n(\textbf{r})\textbf{d}(\textbf{r})dV,
       \label{d4} \end{equation}
where $n(\textbf{r})$ is the microscopic density of dipoles, and
$\textbf{d}(\textbf{r})$ is the dipole moment of a volume, which
contains one atom and has the coordinate $\textbf{r}$. Since
$\int\limits_{V_{+}}\rho(\textbf{r})dV=0$, we may write
$\rho(\textbf{r})= - div \textbf{P}_{mic}(\textbf{r})$,  where
$\textbf{P}_{mic}$ is a microscopic quantity. Similarly to
(\ref{d2}), we obtain
\begin{eqnarray}
\int\limits_{V_{+}}\textbf{r}\rho(\textbf{r})dV
=\int\limits_{V_{+}}\textbf{P}_{mic}(\textbf{r})dV.
       \label{d5} \end{eqnarray}
The relations $\bar{\rho}(\textbf{r})=-div \textbf{P}(\textbf{r})$
and $\bar{\rho}(\textbf{r})=-\overline{ div
\textbf{P}_{mic}(\textbf{r})}=-div
\bar{\textbf{P}}_{mic}(\textbf{r})$ yield $\textbf{P}(\textbf{r})=
\bar{\textbf{P}}_{mic}(\textbf{r})$. Since equalities (\ref{d4}) and
(\ref{d5}) hold for a dielectric of any shape, they yield finally
\begin{equation}
\textbf{P}_{mic}(\textbf{r})=n(\textbf{r})\textbf{d}(\textbf{r}),
\quad
\textbf{P}(\textbf{r})=\overline{n(\textbf{r})\textbf{d}(\textbf{r})}.
       \label{d6} \end{equation}
The possible jump of $\rho(\textbf{r})$ on the surface does not
violate equalities (\ref{d6}), since in the right-hand side of
(\ref{d4}) this jump is smeared over the whole volume. Therefore,
the function $n(\textbf{r})\textbf{d}(\textbf{r})$ is smooth. The
formula $\bar{\rho}(\textbf{r})=-div \textbf{P}(\textbf{r})$ sets
$\textbf{P}(\textbf{r})$ to within $rot \textbf{g}(\textbf{r})$,
where $\textbf{g}(\textbf{r})$ is any function. Therefore, we can
define $\textbf{P}(\textbf{r})$ in the infinite number of ways, but
the chosen way must be consistent with the equation
$\bar{\rho}(\textbf{r})=-div \textbf{P}(\textbf{r})$. For formulae
(\ref{d6}) this holds. According to (\ref{d6}),
$\textbf{P}(\textbf{r})$ is the dipole moment of a unit volume of
the dielectric.

Relation (\ref{d5}) yields also $\textbf{P}(\textbf{r})=\overline{
\textbf{r}\rho(\textbf{r})}$, though
$\textbf{P}_{mic}(\textbf{r})\neq \textbf{r}\rho(\textbf{r})$ (since
$\rho(\textbf{r})$ is a sum of $\delta$-functions, but
$\textbf{P}_{mic}(\textbf{r})=n(\textbf{r})\textbf{d}(\textbf{r})$
is smoothly varied in space).

We now pass to the description of the spontaneous polarization.
Consider an isotropic dielectric characterized in an external
electric field by the dielectric permittivity $\varepsilon$. Let it
contain a macroscopic number of spontaneous dipoles
$\textbf{d}_{s}^{(j)}$, and let it be surrounded by a dielectric
with dielectric permittivity $\varepsilon_{2}$ (without the
intrinsic electromagnetic field). We will write the equations only
for the first dielectric and take the second one into account in
boundary conditions. The spontaneous dipoles are associated with
some average charge density $\bar{\rho}_{s}(\textbf{r})$ and the
polarization $\textbf{P}_{s}$. These dipoles create the electric
field, which polarizes the surrounding atoms (including the atoms,
being the carriers of spontaneous dipoles). This leads to the
appearance of induced charges with average density
$\bar{\rho}_{i}(\textbf{r})$ and the induced polarization
$\textbf{P}_{i}(\textbf{r})$. It is clear that, for the densities
$\bar{\rho}_{s}(\textbf{r})$ and $\bar{\rho}_{i}(\textbf{r}),$ the
total charge is zero:
\begin{equation}
\int\limits_{V_{+}}\bar{\rho}_{s}(\textbf{r})dV=0, \quad
\int\limits_{V_{+}}\bar{\rho}_{i}(\textbf{r})dV=0.
       \label{9} \end{equation}
Similarly to the above analysis, (\ref{9}) yield the formulae
\begin{equation}
\bar{\rho}_{s}(\textbf{r})=-div \textbf{P}_{s}(\textbf{r}), \quad
\bar{\rho}_{i}(\textbf{r})=-div \textbf{P}_{i}(\textbf{r}),
       \label{10} \end{equation}
\begin{equation}
\textbf{P}_{s}(\textbf{r})=\overline{
n_{s}(\textbf{r})\textbf{d}_{s}(\textbf{r})}, \quad
\textbf{P}_{i}(\textbf{r})=\overline{
n_{i}(\textbf{r})\textbf{d}_{i}(\textbf{r})},
       \label{10b} \end{equation}
where $n_{s}(\textbf{r})$ and $n_{i}(\textbf{r})$ are the
microscopic densities of spontaneous and induced dipoles.

Next. The averaging of the microscopic Maxwell equation $div
\textbf{E}_{mic}(\textbf{r})=4\pi\rho(\textbf{r})$ over a physically
infinitesimal volume gives the equation
\begin{equation}
 div \textbf{E}(\textbf{r})=4\pi \bar{\rho}(\textbf{r})=4\pi
 [\bar{\rho}_{s}(\textbf{r})+\bar{\rho}_{i}(\textbf{r})].
       \label{11} \end{equation}
In view of (\ref{10}), (\ref{10b}), we can write (\ref{11}) in the
form
\begin{equation}
 div \textbf{D}(\textbf{r})=0,
       \label{12} \end{equation}
where
\begin{equation}
 \textbf{D}=\textbf{E}+4\pi
 (\textbf{P}_{i}+\textbf{P}_{s}).
       \label{13} \end{equation}
As was mentioned above, the polarization
$\textbf{P}_{i}(\textbf{r})$ is a local response of the medium to
the field $\textbf{E}(\textbf{r})$. Therefore, the relation
\begin{equation}
4\pi
\textbf{P}_{i}(\textbf{r})=(\varepsilon(\textbf{r})-1)\textbf{E}(\textbf{r})
       \label{13b} \end{equation}
should be satisfied analogously to the polarization of a dielectric
by an external field $\textbf{E}_{0}$. Here, $\varepsilon$ is the
usual dielectric permittivity. Indeed, the response is independent
of which force has created the field $\textbf{E}_{0}$ at the given
point of the dielectric. $\textbf{E}_{0}$ can be created by free
charges outside the system or by spontaneous charges inside the
system. In both cases, the value of $\textbf{P}_{i}$ must be the
same. The resulting field is
$\textbf{E}=\textbf{E}_{0}+\textbf{E}_{i}$, where $\textbf{E}_{i}$
is the field created by induced dipoles. Since $\textbf{P}_{i}$ is
identical in both cases, $\textbf{E}$ should be also the same.

It is convenient to introduce the quantity
\begin{equation}
 \textbf{D}_{i}=\textbf{E}+4\pi \textbf{P}_{i}=\varepsilon\textbf{E},
       \label{15} \end{equation}
then Eq. (\ref{11}) takes the form
\begin{equation}
 div \textbf{D}_{i}(\textbf{r})=4\pi \bar{\rho}_{s}(\textbf{r}).
       \label{14} \end{equation}
Thus, one should solve Eqs. (\ref{15}), (\ref{14}) with regard for
the boundary conditions
\begin{equation}
D_{n}=D_{2n}, \quad \textbf{E}_{t}=\textbf{E}_{2t}
       \label{bc} \end{equation}
and the relation
\begin{equation}
 \textbf{E}=-\nabla \varphi-\frac{1}{c}\frac{\partial \textbf{A}}{\partial t},
       \label{16} \end{equation}
which follows from (\ref{6}). In a stationary problem, the magnetic
field is absent: $ \textbf{A}=0, \textbf{B}= rot \textbf{A}=0$. If
spontaneous charges are moving, then a current $\textbf{j}_{s}$ and
a magnetic field arise. The charge moving with a velocity
$\textbf{v}$ creates the potential $\varphi$ and the vector
potential $\textbf{A}=\varphi \textbf{v}/c$ \cite{land2} (in the
immovable reference system). The last relation indicates that the
magnetic field is weak for the processes, which are slow as compared
with the electromagnetic wave.  Therefore, for the stationary and
slow processes, we may set in (\ref{16}) $\partial
\textbf{A}/\partial t = 0$, and Eq. (\ref{14}) takes the form of the
Poisson equation
\begin{equation}
 \triangle\varphi=-4\pi \bar{\rho}_{s}/\varepsilon.
       \label{17} \end{equation}
We neglect the nonuniformity of $\varepsilon$, which is justified
for uniform and weakly nonuniform fields.

Let the spontaneous dipoles
$\textbf{d}_{s}^{(j)}=|q^{(j)}_{0}|\textbf{r}^{(j)}_{0}$
($q^{(j)}_{0}<0$)  be distributed in dielectric. The average density
of effective spontaneous charges reads
\begin{equation}
 \bar{\rho}_{s}=\prec\sum\limits_{j=1}^{N_{s}}[q^{(j)}_{0}\delta(\textbf{r}-\textbf{r}_{j})-q^{(j)}_{0}
 \delta(\textbf{r}-\textbf{r}_{j}-\textbf{r}_{0}^{(j)})]\succ,
       \label{18} \end{equation}
where $N_{s}$ is the number of spontaneous dipoles, $\textbf{r}_{j}$
and $\textbf{r}_{j}+\textbf{r}_{0}^{(j)}$ are the coordinates of the
effective charges $q^{(j)}_{0}$, $-q^{(j)}_{0}$ of the $j$-th
dipole, and $\prec f\succ \equiv \bar{f}$. The solution of Eqs.
(\ref{17}), (\ref{18}) is known:
\begin{equation}
 \varphi(\textbf{r})=\prec\sum\limits_{j=1}^{N_{s}}\left [\frac{q^{(j)}_{0}}{\varepsilon|\textbf{r}-\textbf{r}_{j}|}-
 \frac{q^{(j)}_{0}}{\varepsilon|\textbf{r}-\textbf{r}_{j}-\textbf{r}_{0}^{(j)}|}
 \right ]\succ.
       \label{19} \end{equation}
For the points $\textbf{r}$ far from the spontaneous dipoles
($|\textbf{r}-\textbf{r}_{j}|\gg r_{0}^{(j)}$), we can make
expansion in $\textbf{r}_{0}^{(j)}$. As a result, we have
\begin{eqnarray}
&&\varphi(\textbf{r})= \prec
\sum\limits_{j=1}^{N_{s}}\frac{\textbf{d}_{s}^{(j)}\cdot(\textbf{r}-\textbf{r}_{j})}{\varepsilon|\textbf{r}-\textbf{r}_{j}|^{3}}\succ
=\nonumber \\&&=
\prec\int\limits_{V}d\acute{\textbf{r}}\frac{n_{s}(\acute{\textbf{r}})\textbf{d}_{s}(\acute{\textbf{r}})\cdot
(\textbf{r}-\acute{\textbf{r}})}{\varepsilon|\textbf{r}-\acute{\textbf{r}}|^{3}}\succ
= \label{20} \\ &&=
\int\limits_{V}d\acute{\textbf{r}}\frac{\overline{
n_{s}(\acute{\textbf{r}})\textbf{d}_{s}(\acute{\textbf{r}})}
(\textbf{r}-\acute{\textbf{r}})}{\varepsilon|\textbf{r}-\acute{\textbf{r}}|^{3}}=
\int\limits_{V}d\acute{\textbf{r}}\frac{\textbf{P}_{s}(\acute{\textbf{r}})\cdot
(\textbf{r}-\acute{\textbf{r}})}{\varepsilon|\textbf{r}-\acute{\textbf{r}}|^{3}},
       \nonumber \end{eqnarray}
where $V$ is the volume of the dielectric.

Formula (\ref{20}) without $\varepsilon$ in the denominator is well
known \cite{tamm}. The quantity $\varepsilon$ in (\ref{20}) takes
the polarization $\textbf{P}_{i}$ into account. Thus, taking
$\textbf{P}_{i}$ into consideration leads to the replacements
$\textbf{d}_{s}\rightarrow \textbf{d}_{s}/\varepsilon$,
$\textbf{P}_{s}\rightarrow \textbf{P}_{s}/\varepsilon$ in
(\ref{20}). This change has a simple physical meaning: it shows that
the medium weakens the field of a dipole by $\varepsilon$ times. In
view of this reasoning, formula (\ref{20}) was intuitively guessed
in \cite{mt2010,mt2011}.

For the uniform spontaneous polarization, we have $\textbf{P}_{s}
=(N_{s}/V)\textbf{d}_{s}=\bar{n}_{s}\textbf{d}_{s}$ and
\begin{equation}
\varphi(\textbf{r})=
\int\limits_{V}d\acute{\textbf{r}}\frac{\textbf{P}_{s}\cdot
(\textbf{r}-\acute{\textbf{r}})}{\varepsilon|\textbf{r}-\acute{\textbf{r}}|^{3}}=
-\textbf{P}_{s}\nabla_{\textbf{r}}\int\limits_{V}\frac{d\acute{\textbf{r}}}{\varepsilon|\textbf{r}-\acute{\textbf{r}}|}.
       \label{21} \end{equation}
 With regard for the formula \cite{korn}
\begin{eqnarray}
\nabla(\textbf{F}\textbf{G})&=&(\textbf{F}\nabla)\textbf{G}+(\textbf{G}\nabla)\textbf{F}+
\textbf{F}\times(\nabla \times\textbf{G})+\nonumber
\\&+&\textbf{G}\times(\nabla \times\textbf{F}),
       \label{22} \end{eqnarray}
we find  from (\ref{21}):
\begin{equation}
\textbf{E}(\textbf{r})=-\nabla_{\textbf{r}}\varphi(\textbf{r})=(\textbf{P}_{s}\nabla_{\textbf{r}})\nabla_{\textbf{r}}f(\textbf{r}),
       \label{23} \end{equation}
\begin{equation}
f(\textbf{r})=\int\limits_{V}\frac{d\acute{\textbf{r}}}{\varepsilon|\textbf{r}-\acute{\textbf{r}}|}.
       \nonumber \end{equation}

Formulae (\ref{20}) and (\ref{23}) give the solution  for the field
$\textbf{E}(\textbf{r})$ created by a spontaneously polarized
isotropic dielectric of volume $V$,  surrounded by va\-cuum. It is
seen from (\ref{23}) that $\textbf{E}(\textbf{r})=const \cdot
\textbf{P}_{s}$ only in particular cases, for example, if
$f(\textbf{r})=b_{0}+\textbf{b}_{1}\textbf{r} +b_{2}\textbf{r}^{2}$
or if $\textbf{P}_{s}=P_{s}\textbf{i}_{z}$,
$f(\textbf{r})=b_{0}+b_{1}z+b_{2}z^{2}$. The symmetry-based
reasoning implies that, for a finite $V,$ the first case with
$\textbf{b}_{1}=0$ is possible only for the dielectrics with shape
of a ball (the case with $\textbf{b}_{1}\neq 0$ corresponds probably
to an ellipsoid), but the second case is impossible for a
three-dimensional system. That is, $\textbf{E}(\textbf{r})$ is not
codirected with $\textbf{P}_{s}$ in the general case. Moreover,
$\textbf{E}$ can be nonuniform, when $\textbf{P}_{s}$ is uniform. We
have verified these properties for a cylindrical dielectric with
$\textbf{P}_{s}=P_{s}\textbf{i}_{z}$ (see also Sec. 6.1). These
properties suggest that, for an isotropic spontaneously polarized
dielectric, the relationship between the electric induction
\begin{equation}
 \textbf{D}=\textbf{D}_{i}+4\pi\textbf{P}_{s}=\varepsilon\textbf{E}+4\pi\textbf{P}_{s}
       \label{24} \end{equation}
and the strength $\textbf{E}$ takes generally the tensor form:
\begin{equation}
 D_{j}=\sum_{l}\tilde{\varepsilon}_{jl}E_{l}, \quad \tilde{\varepsilon}_{jl}
 =\delta_{jl}\varepsilon+\zeta_{jl}.
       \label{24main} \end{equation}
According to Eqs. (\ref{20}) and (\ref{23}), the field
$\textbf{E}(\textbf{r})$ is connected with
$\textbf{P}_{s}(\textbf{r})$ by the relation
$\textbf{E}(\textbf{r})=\hat{\varsigma}(\textbf{r})\textbf{P}_{s}(\acute{\textbf{r}})$,
where $\hat{\varsigma}(\textbf{r})$ is a linear integro-differential
operator. In this case, we can write
$E_{j}(\textbf{r})=\sum_{l}\varsigma_{jl}(\textbf{r})P_{s,l}(\textbf{r})$
and $\zeta_{jl}(\textbf{r})=4\pi\varsigma_{jl}^{-1}(\textbf{r})$,
where $\varsigma_{jl}(\textbf{r})$ is the matrix of eigenvalues of
the operator $\hat{\varsigma}(\textbf{r})$. We note that the tensor
$\zeta_{jl}$ characterizes the spontaneous polarization and is
physically different  from the tensor of dielectric permittivity
$\varepsilon_{jl}$. We also note that $\tilde{\varepsilon}_{jl}$ is
not the tensor of dielectric permittivity as well, though it
contains $\varepsilon$. Indeed, the quantity $\varepsilon_{jl}$
characterizes the response $\textbf{P}_{i}$ of the system to an
external field $\textbf{E}$. On the contrary, the quantity
$\zeta_{jl}$ allows one to determine the field
$\textbf{E}(\textbf{r})$ created by the polarization
$\textbf{P}_{s}(\textbf{r})$. Moreover, $\varepsilon_{jl}$ is a
local quantity. However, $\zeta_{jl}$ is a nonlocal quantity, since
its value is determined by the distribution of the spontaneous
polarization in the whole volume of the dielectric and by the
boundary conditions. For an anisotropic dielectric, polarized by an
external electric field, the principal values of the tensor
$\varepsilon_{jl}$ must be $\geq 1$ \cite{land8}. However, for
$\zeta_{jl}$ and $\tilde{\varepsilon}_{jl}$, such restriction is
absent. The tensor character of $\zeta_{jl}$ is a consequence of the
fact that the field $\textbf{E}(\textbf{r})$ (\ref{23}) created by
spontaneous dipoles is not codirected with
$\textbf{P}_{s}(\textbf{r})$ in the general case. If
$\textbf{E}(\textbf{r})$ and $\textbf{P}_{s}(\textbf{r})$ are
codirected, then $\zeta(\textbf{r})$ is a scalar:
$\zeta_{jl}(\textbf{r})=\delta_{jl}\zeta(\textbf{r})$,
$4\pi\textbf{P}_{s}=\zeta(\textbf{r})\textbf{E}(\textbf{r})$.

We remark that the nonlocal connection between the intrinsic field
$\textbf{E}(\textbf{r})$ and $\textbf{P}(\textbf{r})$ was derived
previously for ferroelectrics
\cite{binder1979,moroz2007,moroz2009,moroz2016}. In this case, the
solutions \cite{binder1979,moroz2009,moroz2016} have the local form
$\textbf{E}(\textbf{r})=const \textbf{P}(\textbf{r})$ (the
nonlocality is hidden due to the one-dimensional geometry and/or the
uniformity of $\textbf{P}$). Ferroelectrics were considered
\cite{binder1979,moroz2007,moroz2009,moroz2016} without the
separation of the total polarization $\textbf{P}$ into
$\textbf{P}_{s}$ and $\textbf{P}_{i}$.

In ferroelectrics, the distribution of the polarization is regulated
by the minimum of the thermodynamic potential $\Phi$
\cite{land8,strukov,moroz2009,ginzburg2001,ginzburg1949}. Therefore,
it should consider namely the total polarization
$\textbf{P}(\textbf{r})$ to be known. In such case, the above
formulae are applicable, if we change $\bar{\rho}_{s}\rightarrow
\bar{\rho}$, $\textbf{P}_{s}(\textbf{r})\rightarrow
\textbf{P}(\textbf{r})$, and $\bar{\rho}_{i},
\textbf{P}_{i}(\textbf{r})\rightarrow 0$. This is equivalent to one
change: $\varepsilon\rightarrow 1$. For the piezoelectrics, one can
represent $\textbf{P}(\textbf{r})$ in the form of the deformational
and induced parts (similarly to Eq. (\ref{f-13}) below, but in the
tensor form) \cite{land8,born}. In this case, the second term
affects the first one by means of the equations of elastic
equilibrium. For the nonpolar dielectrics, the influence of
$\textbf{P}_{i}$ on $\textbf{P}_{s}$ is negligible (see the
following section). Such difference is apparently due to the fact
that the force, acting on the volume element, is $F\sim E$ for a
piezoelectric and $F\sim E^{2}$ for a nonpolar dielectric (in this
case, the field $E$ is weak; here, we talk about the electric part
of $\textbf{F}$).

Thus, in order to find the field $\textbf{E}(\textbf{r})$ in a
spontaneously polarized nonpolar isotropic dielectric, we need to
solve Eqs. (\ref{15}), (\ref{14}) with the boundary conditions
(\ref{bc}), in which $\textbf{D}$ is given by the formula
(\ref{24}). If the dielectric is surrounded by vacuum, then Eq.
(\ref{20}) gives the exact solution. In this case, the boun\-dary
conditions (\ref{bc}) are satisfied automatically.

Let us turn to the Maxwell equations (\ref{5})--(\ref{8}). We noted
above that in the presence of the spontaneous polarization one
should solve Eq. (\ref{14}) (with the replacement
$\bar{\rho}_{s}(\textbf{r})\rightarrow
\bar{\rho}_{s}(\textbf{r})+\rho_{f}(\textbf{r})$, if foreign charges
are present) instead of Eq. (\ref{5}). In this case, Eq. (\ref{14})
is simply another form of Eq. (\ref{5}). Equations (\ref{6}) and
(\ref{7}) remain the same. Equation (\ref{8}) can be verified
similarly to the analysis in \cite{land8} (Chapt. IX, $\S 75$). We
conclude that Eq. (\ref{8}) is valid. In this case, $\textbf{D}$ is
given by formula (\ref{24}). However, while solving Eq. (\ref{8}),
we should separate $\textbf{P}_{s}$ from $\textbf{D}$ by means of
the replacements $\textbf{D}\rightarrow \textbf{D}_{i}$ and
$\textbf{j}_{f}\rightarrow \textbf{j}_{f}+\bar{\textbf{j}}_{s}$,
where $\bar{\textbf{j}}_{s}=\overline{\rho_{s}\textbf{v}_{s}}$ is
the current related to spontaneous dipoles. Thus, the Maxwell
equations (\ref{5})--(\ref{8}) conserve formally their validity.
However, while solving them, it is necessary to separate
$\textbf{P}_{s}$ from $\textbf{D}$, since the spontaneous dipoles
are the source of the field. In addition, $\textbf{D}$ is given by
formulae (\ref{24}),  (\ref{24main}) instead of (\ref{1}).

In  experiments, the potential $\varphi(\textbf{r})$ is measured
du\-ring a macroscopic time. For such time interval, a configuration
of spontaneous dipoles change a huge number of times. Therefore, we
need  to average $\varphi(\textbf{r})$ over the time. According to
Gibbs \cite{gibbs}, the average over the time can be replaced by the
average over the ensemble. Quantum statistics gives the following
formula for the average over the ensemble \cite{bog}:
\begin{equation}
\langle\hat{\varphi}\rangle=\int d\Omega
Z^{-1}\sum\limits_{n}e^{-E_{n}/k_{B}T}\Psi^{*}_{n}\hat{\varphi}
\Psi_{n},
       \label{31} \end{equation}
where  $Z=\sum_{n}e^{-E_{n}/k_{B}T}$,  $\hat{\varphi}$ is given by
(\ref{20}), $E_{n}$ is the energy of the system in the $n$-th state,
and $\{\Psi_{n}\}$ is the complete collection of wave functions of
the system. The operator $\hat{\varphi}$ and the functions
$\Psi_{n}$ should be written in terms of the coordinates of the
nucleus and electrons of each atom. Thus, the polarization
$\textbf{P}_{s}(\acute{\textbf{r}})$ in (\ref{20}) should be
additionally averaged statistically. Then
$\textbf{P}_{s}(\acute{\textbf{r}})=\langle\overline{n_{s}(\acute{\textbf{r}})\textbf{d}_{s}(\acute{\textbf{r}})}\rangle=
\overline{\langle
n_{s}(\acute{\textbf{r}})\textbf{d}_{s}(\acute{\textbf{r}})\rangle
}$.

\section{Mechanical response of a dielectric to an external electric field}
In Section 2 we assumed that the quantity $\textbf{P}_{s}$ is fixed
and can be obtained with regard for the mechanism of spontaneous
polarization. The fixity of $\textbf{P}_{s}$ means that the field
$\textbf{E}(\textbf{r})$ created by dipoles of the medium has no
effect on $\textbf{P}_{s}(\textbf{r})$. We now find the influence of
the field $\textbf{E}$ on the value of $\textbf{P}_{s}$, which
allows us to write the complete system of equations for
$\textbf{P}(\textbf{r})$ and $\textbf{E}(\textbf{r})$. The
polarization $\textbf{P}_{s}(\textbf{r})$ arises at $\nabla n\neq
0$. Therefore, we should determine the influence of the field
$\textbf{E}$ on the density $n(\textbf{r})$. The motion of the ideal
fluid is described by the Euler equation
  \begin{equation}
\rho D\textbf{v}/Dt\equiv \rho \partial\textbf{v}/\partial t
+\rho(\textbf{v}\nabla)\textbf{v}  = \textbf{F}
  \label{f-10} \end{equation}
and by the equation of continuity
 \begin{equation}
\partial\rho/\partial t + div (\rho\textbf{v})  =0,
  \label{f-11} \end{equation}
where
 \begin{equation}
\textbf{F}=-\nabla p+\textbf{F}^{nm},
  \label{f-11b} \end{equation}
$\textbf{F}^{nm}$ is a non-mechanical (usually  external) force per
unit volume, $\rho =mn$, and $m$ is the mass of an atom of the
dielectric. In the presence of the electric field, $\textbf{F}^{nm}$
is approximately the force, with which the field $\textbf{E}$ acts
on the dipole moment of a unit volume. Since the force acting on the
charge $q$ is $q\textbf{E}+[\textbf{v}\times \textbf{H}]q/c$, we
find, by neglecting the magnetic field,
\begin{equation}
\textbf{F}^{nm}\approx (\textbf{P}\nabla)\textbf{E}, \quad
\textbf{E}(\textbf{r})=\textbf{E}^{own}(\textbf{r})+\textbf{E}^{ext}(\textbf{r}).
  \label{f-12} \end{equation}
Here, $\textbf{E}^{own}$ is the field induced by the dipoles of the
dielectric, and $\textbf{E}^{ext}$ is the external field. According
to the above analysis, we get
\begin{equation}
\textbf{P}=\textbf{P}_{s}+\textbf{P}_{i}= \xi\nabla
n(\textbf{r})+\frac{\varepsilon -1}{4\pi}\textbf{E}.
  \label{f-13} \end{equation}
Therefore,
\begin{equation}
\textbf{F}^{nm}\approx \xi(\nabla n \cdot \nabla)\textbf{E}+
\frac{\varepsilon -1}{8\pi}\nabla E^{2}.
  \label{f-14} \end{equation}
Without an external field ($\textbf{E}^{ext}=0$), we have $
|F^{nm}|\ll |\nabla p|$. Indeed, under the spontaneous polarization,
the relation $E\sim P\sim P_{s}$ usually holds (see Appendix).
Setting $E=P_{s}$ in (\ref{f-14}) and using formula (\ref{0-1}) with
$\xi= (7/3)d_{0}\bar{r}_{0}(n(\textbf{r})/n_{0})^{2}$, we get $
|F^{nm}|/|\nabla p|\sim 10^{-16}\bar{r}^{2}_{0}|(\triangle
n)/n+2((\nabla n)/n)^{2}|$ for He-II (for other dielectrics,
$|F^{nm}|/|\nabla p|$ can be different by several orders of
magnitude). This is a negligible value, since $
\bar{r}^{2}_{0}|(\triangle n)/n+2((\nabla n)/n)^{2}|\lsim 1$ for
real systems. From the physical point of view, the smallness of
$|F^{nm}|/|\nabla p|$ is related to the small value of the mutual
polarization of two nonpolar atoms (e.g., the electron shell of a He
II atom is elongated due to the interaction with adjacent atoms by
$\sim 10^{-5}\mbox{\AA}$) and to that $F^{nm}\sim E^{2}$. Therefore,
we may neglect the term $\textbf{F}^{nm}$ in Eqs. (\ref{f-10}),
(\ref{f-11b}). Hence, for $\textbf{E}^{ext}=0,$ the influence of the
field $\textbf{E}$ on the velocity of an element of the fluid and on
the density can be neglected. Therefore, we can consider the
polarization $\textbf{P}_{s}$ in Eq. (\ref{20}) to be a
\textit{fixed} quantity depending on $\nabla n$ and independent of
$\textbf{E}$ (as was assumed in Section 2).

The more accurate derivation of the expression for a force is
complicated and gives Helmholtz's formula (see \cite{land8}, Chap.
II):
\begin{equation}
\textbf{F}^{nm}= \frac{1}{8\pi}\nabla\left [E^{2}\rho\left
(\frac{\partial\varepsilon}{\partial \rho}\right )_{T}\right
]-\frac{E^{2}}{8\pi}\nabla\varepsilon.
  \label{f-15b} \end{equation}
This formula is obtained for the polarization by the external field
($\textbf{D}=\varepsilon \textbf{E}$) and does not involve the
spontaneous polarization $\textbf{P}_{s}$.

For a spontaneously polarized dielectric, we should consider also
the polarization $\textbf{P}_{s}$, since
\begin{equation}
 \textbf{D}=\varepsilon \textbf{E}+4\pi\textbf{P}_{s}
       \label{a-0} \end{equation}
(see Eq. (\ref{24})). Let us determine the force $\textbf{F}$ with
regard for the contribution of $\textbf{P}_{s}$. We will apply the
analysis made in \cite{land8} and will present only the changes
caused by $\textbf{P}_{s}$. As for the details, see \cite{land8},
$\S 15, 16$. We are based on the formulae, obtained for a liquid
dielectric \cite{land8}:
\begin{equation}
 F_{i}=\sum\limits_{k}\frac{\partial \sigma_{ik}}{\partial x_{k}},
       \label{a-1} \end{equation}
\begin{equation}
 \sigma_{ik}=\left [\tilde{F}-\rho \frac{\partial \tilde{F}}{\partial \rho}|_{\textbf{E},T}  \right ]\delta_{ik}+\frac{E_{i}D_{k}}{4\pi},
       \label{a-2} \end{equation}
\begin{equation}
d\tilde{F}=-SdT+\zeta d\rho-\frac{1}{4\pi}\textbf{D}d \textbf{E}.
       \label{a-3} \end{equation}
Here, $\sigma_{ik}$ is the stress tensor, $\tilde{F}$ is the free
energy per unit volume, and $\zeta$ is the chemical potential per
unit mass.

Since $\textbf{P}_{s}$ is independent of $\textbf{E}$, relations
(\ref{a-0}) and (\ref{a-3}) yield
\begin{equation}
\tilde{F}=\tilde{F}_{0}(\rho,T)-\frac{\varepsilon
E^{2}}{8\pi}-\textbf{P}_{s}\textbf{E}.
       \label{a-4} \end{equation}
The account for $\textbf{P}_{s}$ according to  (\ref{a-0}),
(\ref{a-4}) gives the following addition to $\sigma_{ik}$:
\begin{equation}
 \sigma^{s}_{ik}=\left [-\textbf{P}_{s}\textbf{E}+\left (\rho \frac{\partial \textbf{P}_{s}}{\partial \rho}|_{T}\right )\textbf{E}
 \right ]\delta_{ik}+E_{i}P_{s,k}.
       \label{a-5} \end{equation}
However, the complete tensor  $\sigma_{ik}$ for such solution has no
symmetry $\sigma_{ik}=\sigma_{ki}$, which should hold
\cite{land8,land7}. In order to restore this symmetry, we take into
account that the spontaneous polarization
$\textbf{P}_{s}(\textbf{r})$ separates a local axis in the medium
and leads to the anisotropy (though the distribution of atoms is
isotropic). The anisotropy is also indicated by formula
(\ref{24main}). By means of a rotation of the coordinate system, the
tensor $\tilde{\varepsilon}_{jl}$ can be diagonalized, which
separates also definite axes. To account for such anisotropy, we can
use the reasoning for the crystal axes  (see \cite{land8}, $\S 16$).
The result consists in the requirement to make change
\begin{equation}
 E_{i}D_{k}\rightarrow
(E_{i}D_{k}+E_{k}D_{i})/2
       \label{a-ee} \end{equation}
in (\ref{a-2}). Then relation  (\ref{a-2}) becomes
\begin{equation}
 \sigma_{ik}=\left [\tilde{F}-\rho \frac{\partial \tilde{F}}{\partial \rho}|_{\textbf{E},T}  \right ]\delta_{ik}+\frac{E_{i}D_{k}+E_{k}D_{i}}{8\pi},
       \label{a-2b} \end{equation}
and $\sigma^{s}_{ik}$ takes the form
\begin{equation}
 \sigma^{s}_{ik}=\left [-\textbf{P}_{s}\textbf{E}+\left (\rho \frac{\partial \textbf{P}_{s}}{\partial \rho}|_{T}\right )\textbf{E}
 \right ]\delta_{ik}+\frac{E_{i}P_{s,k}+E_{k}P_{s,i}}{2}.
       \label{a-5b} \end{equation}
According to Introduction, $\textbf{P}_{s}(\textbf{r}) = const \cdot
n^{a}\nabla n$. Therefore, we set $\rho \frac{\partial
\textbf{P}_{s}}{\partial \rho}=a \textbf{P}_{s}$. Then
\begin{equation}
 \sigma^{s}_{ik}=(a-1)\textbf{P}_{s}\textbf{E}\delta_{ik}+ \frac{E_{i}P_{s,k}+E_{k}P_{s,i}}{2}.
       \label{a-6} \end{equation}
By (\ref{a-1}), this gives the additional force
\begin{eqnarray}
 \textbf{F}_{s}&=&(a-1)\nabla(\textbf{P}_{s}\textbf{E})+\frac{1}{2}\textbf{E}\cdot div
 \textbf{P}_{s}+ \nonumber \\ &+&
 \frac{1}{2}(\textbf{P}_{s}\nabla)\textbf{E}+ \frac{1}{2}\textbf{P}_{s}\cdot div \textbf{E}+\frac{1}{2}(\textbf{E}\nabla)\textbf{P}_{s}.
       \label{a-7} \end{eqnarray}
By assuming $\textbf{D}=\varepsilon \textbf{E},$ the formula
\begin{eqnarray}
 F_{j}&=&\frac{\partial }{\partial x_{j}}\left [-p+ \frac{E^{2}}{8\pi}
 \rho\frac{\partial \varepsilon}{\partial \rho}|_{T}\right ]-\frac{E^{2}}{8\pi}\frac{\partial \varepsilon}{\partial x_{j}}+
 \nonumber \\ &+&\frac{1}{4\pi}\left [-\frac{\varepsilon}{2}\frac{\partial E^{2}}{\partial x_{j}}
 +\sum\limits_{k}\frac{\partial}{\partial x_{k}}(\varepsilon E_{j}E_{k}) \right ]
       \label{a-8} \end{eqnarray}
was obtained in \cite{land8}. Further, the property $div
\textbf{D}=\sum_{k}\frac{\partial}{\partial x_{k}} (\varepsilon
E_{k})=0$ was used in \cite{land8}. This property yields
$-\frac{\varepsilon}{2}\frac{\partial E^{2}}{\partial x_{j}}
 +\sum_{k}\frac{\partial}{\partial x_{k}}(\varepsilon
 E_{j}E_{k})=0$ in (\ref{a-8}). In our case, relation (\ref{a-0}) leads to
the formula $div \textbf{D}=\sum_{k}\frac{\partial}{\partial x_{k}}
(\varepsilon E_{k})+4\pi \sum_{k}\frac{\partial}{\partial x_{k}}
(P_{s,k})=0$. Therefore,
\begin{equation}
 -\frac{\varepsilon}{2}\frac{\partial E^{2}}{\partial x_{j}}
 +\sum\limits_{k}\frac{\partial}{\partial x_{k}}(\varepsilon
 E_{j}E_{k})=-4\pi E_{j}\cdot div \textbf{P}_{s}.
       \label{a-9} \end{equation}
Formulae (\ref{a-7}), (\ref{a-8}), and (\ref{a-9}) yield the
solution:
\begin{eqnarray}
 \textbf{F}&=&\nabla\left [-p+ \frac{E^{2}}{8\pi}
 \rho\frac{\partial \varepsilon}{\partial \rho}|_{T}\right ]-\frac{E^{2}}{8\pi}\nabla \varepsilon+
 \nonumber \\ &+& (a-1)\nabla(\textbf{P}_{s}\textbf{E})-\frac{1}{2}\textbf{E}\cdot div \textbf{P}_{s}+
\frac{1}{2}\textbf{P}_{s}\cdot div \textbf{E}+\nonumber \\ &+&
\frac{1}{2}(\textbf{P}_{s}\nabla)\textbf{E}+\frac{1}{2}(\textbf{E}\nabla)\textbf{P}_{s}.
       \label{a-10} \end{eqnarray}
Using relations (\ref{22}), $ rot \textbf{E}=0$, $ rot
\textbf{P}_{s}=0$ (since  $\textbf{P}_{s}(\textbf{r}) = const \nabla
n^{a+1}$), and  \cite{korn}
\begin{eqnarray}
rot
(\textbf{F}\times\textbf{G})&=&(\textbf{G}\nabla)\textbf{F}-(\textbf{F}\nabla)\textbf{G}+
\textbf{F}\cdot div\textbf{G}-\nonumber
\\&-&\textbf{G}\cdot div\textbf{F},
       \label{a-11} \end{eqnarray}
we obtain finally the formula for a force $\textbf{F}$ acting on a
volume element $dV$ of a spontaneously polarized liquid dielectric:
\begin{eqnarray}
 \textbf{F}&=&\nabla\left [-p+ \frac{E^{2}}{8\pi}
 \rho\frac{\partial \varepsilon}{\partial \rho}|_{T}\right ]-\frac{E^{2}}{8\pi}\nabla \varepsilon+
\label{a-12} \\ &+&
(a-1)\nabla(\textbf{P}_{s}\textbf{E})+(\textbf{P}_{s}\nabla)\textbf{E}+\frac{1}{2}rot
(\textbf{P}_{s}\times\textbf{E}).
      \nonumber \end{eqnarray}
Three last terms arose due to the account for the spontaneous
polarization.

Let us use the Clausius--Mossotti formula \cite{tamm}
$\frac{\varepsilon -1}{\varepsilon +2}=\frac{4\pi}{3}n\beta$, which
holds for nonpolar fluids and gases (here, $\beta$ is the
polarizability of a molecule). For the gases and some fluids
including He II \cite{eselson1978}, $\varepsilon$ is close to 1.
Then $\varepsilon -1\approx 4\pi n\beta$ and $\rho
\partial\varepsilon/\partial\rho \approx \varepsilon-1.$ Instead of (\ref{a-12}),
we obtain
\begin{eqnarray}
\textbf{F}&=& -\nabla p + \frac{\varepsilon -1}{8\pi}\nabla
E^{2}+(\textbf{P}_{s}\nabla)\textbf{E} +\nonumber \\ &+&
(a-1)\nabla(\textbf{P}_{s}\textbf{E}) +\frac{1}{2}rot
(\textbf{P}_{s}\times\textbf{E}).
  \label{f-15} \end{eqnarray}
This formula differs from (\ref{f-14}) by two additional last terms.
In this case, the term $(a-1)\nabla(\textbf{P}_{s}\textbf{E})$
disappears at $a=1$. Interestingly, the solution $\xi\approx
7.5(7/3)d_{0}\bar{r}_{0}(n(\textbf{r})/n_{0})$ \cite{mt2010}
corresponds namely to $a=1$. Such value of $a$ is a scaling property
\cite{mt2010}. Therefore, it should hold for any nonpolar fluid.

\section{Conclusion}
We have proposed a method of description of a nonpolar isotropic
spontaneously polarized dielectric that takes into account the
induced polarization $\textbf{P}_{i}(\textbf{r})$, which accompanies
the bare spontaneous polarization $\textbf{P}_{s}(\textbf{r})$. In
Appendix we have also obtained (as examples) several solutions for
the  electric field in a spontaneously polarized isotropic
dielectric under different boundary conditions.

\vskip3mm \textit{The author is grateful to A.\,S. Rybalko for the
detailed discussion of the experiments \cite{ryb2004}. I also thank
A. Morozovska for the discussion of the problem. The present work
was partially supported by the National Academy of Sciences of
Ukraine (project No.~0116U003191).}

\section{Appendix. Examples of solutions}
 The below-presented examples can be useful for the
comprehension of properties of the field $\textbf{E}$ inside of a
resonator in experiments similar to
\cite{ryb2005,ryb2004,chag2016,yayama2018,chag2017}. In those
experiments, the boundary conditions (BCs) are complicated. We will
consider more simple BCs and will see how the shape of a resonator
and the presence of an external metallic shell affect the field
$\textbf{E}$ (in expe\-riments \cite{ryb2004} a resonator was placed
in metallic shell for the protection against external electric
signals).

\subsection{Spontaneously polarized plate in vacuum}
Consider a dielectric plate, which occupies the space region $x,y
\in [-L/2, L/2], z\in [-H/2,H/2]$. Let the plate have the uniform
total polarization $\textbf{P}=P\textbf{i}_{z}$ (we suppose that
$\textbf{P}$ is known). The plate is surrounded by vacuum.

The knowledge of the solution for the field $\textbf{E}$ in this
system is essential for the experiments like
\cite{ryb2005,ryb2004,chag2016,yayama2018,chag2017} and for the
ferroelectrics. In many works (including well-known ones
\cite{strukov,ginzburg1949}), the field in such system was described
by the formulae
\begin{equation}
 \textbf{E}^{(int)}=-4\pi \textbf{P}, \quad \textbf{E}^{(ext)}=0,
       \label{11-1} \end{equation}
where $\textbf{E}^{(int)}$ and $\textbf{E}^{(ext)}$ are the fields
inside and outside of the plate, respectively. However, the solution
of this problem is different.

The problem can be solved accurately, if we start from Eq.
(\ref{11}), where $\bar{\rho}(\textbf{r})=0$ in the space outside of
a dielectric. Equation (\ref{11}) is true in the whole space.
Therefore, BCs (\ref{bc}) are satisfied automatically. The solution
of Eq. (\ref{11}) is given by formula (\ref{21}), where $V$ is the
volume of a dielectric, and we should set $\varepsilon=1$,
$\textbf{P}_{s}=\textbf{P}$. In this case, we obtain
\begin{eqnarray}
&&\varphi(\textbf{r})=P
\int\limits_{-L/2}^{L/2}d\acute{x}\int\limits_{-L/2}^{L/2}d\acute{y}\cdot
\nonumber \\ &&\cdot \left
(\frac{1}{\sqrt{(H/2-z)^{2}+(x-\acute{x})^{2}+(y-\acute{y})^{2}}}\right.-\nonumber
\\ &&-\left.\frac{1}{\sqrt{(H/2+z)^{2}+(x-\acute{x})^{2}+(y-\acute{y})^{2}}}
\right ).
       \label{11-2} \end{eqnarray}
\begin{figure}
\vskip1mm
\includegraphics[width=\column]{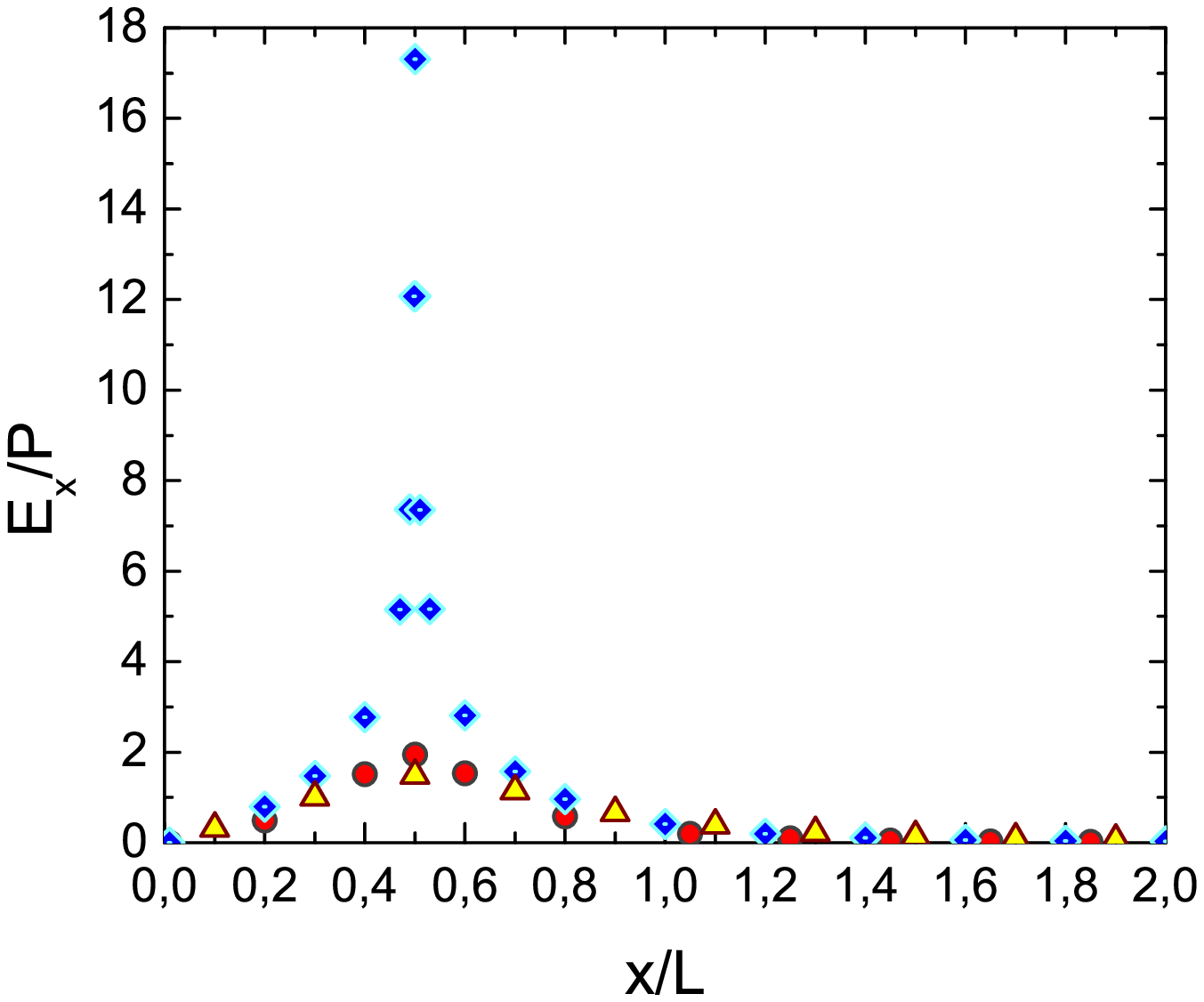}
\vskip-3mm\caption{[Color online] $E_{x}(x)$ for a dielectric plate
with $L=2H$ at $y=0$ and 1) $z=H/4$ (circles),  2) $z=H/2$ (rhombs),
3) $z=H$ (triangles). It is clear from the symmetry that the same
curves correspond to $E_{y}(y)$  for $x=0$ and $z=H/4; H/2; H$. The
lateral surface of the plate corresponds to $x/L=0.5$. }
\end{figure}

This solution has the properties $\varphi(x,y,-z)=-\varphi(x,y,z)$,
$E_{x}(x,y,-z)=-E_{x}(x,y,z)$, $E_{y}(x,y,-z)=-E_{y}(x,y,z)$, and
$E_{z}(x,y,-z)=E_{z}(x,y,z)$. We have found the potential
(\ref{11-2}) and the field $\textbf{E}(\textbf{r})$ numerically, see
Figs. 1, 2 (the field near the plate edge is found approximately;
the field $E_{y}(\textbf{r})$ is very weak at $y=0$ and is not shown
in the figures). It is clear from the symmetry of the problem that
$E_{x}(\textbf{r})$ coincides with $E_{y}(\textbf{r}^{\prime})$,
where $\textbf{r}^{\prime}$ is obtained by the rotation of
$\textbf{r}$ around the axis $z$ by $\pm 90^{\circ}$. As is seen
from the plots, the field $\textbf{E}(\textbf{r})$ is nonuniform and
has a complicated structure. Inside of the dielectric,
$\textbf{E}\approx -2\pi \textbf{P}=-2\pi P\textbf{i}_{z}$. More
accurately, $E_{x}, E_{y}\neq 0$; therefore, the dependence
$\textbf{D}(\textbf{E})$ has a tensor form
$D_{j}=\sum_{l}\tilde{\varepsilon}_{jl}E_{l}$. Outside of the
dielectric, the field $\textbf{E}(\textbf{r})$ is rather strong near
the dielectric and decreases with distance. The direct checking
indicates that BCs (\ref{bc}) hold on all surfaces of the plate.

To verify the solution, we will find it by another method. We start
from Eq. (\ref{11}) in the form
\begin{equation}
 \triangle\varphi=-4\pi \bar{\rho}.
       \label{11-3} \end{equation}
The equation $\bar{\rho}(\textbf{r})=-div \textbf{P}(\textbf{r})$
implies that the polarization $\textbf{P}=P\textbf{i}_{z}$
($P=const$) is equivalent to the charges on the lower and upper
faces of a dielectric with the surface densities $\sigma(z=-H/2)=-P$
and $\sigma(z=H/2)=P,$ respectively. We assume that there are
$N_{q}$ point charges $q$ on the lower face and $N_{q}$ charges
equal to $-q$ on the upper face. In this case, $N_{q}q/L^{2}=-P$,
$N_{q}\rightarrow \infty$. Then
\begin{equation}
 \bar{\rho}=\sum\limits_{j=1}^{N_{q}}q\delta(x-x_{j})\delta(y-y_{j})\left
[\delta\left (z+\frac{H}{2}\right )-\delta\left (z-\frac{H}{2}\right
)\right ],
       \label{11-4} \end{equation}
where $x_{j}, y_{j} \in [-L/2,L/2]$. We consider the space to be a
closed cube with the volume $V_{s}=\Lambda^{3}$ ($\Lambda\gg L$) and
expand $\bar{\rho}$ and $\varphi$ in Fourier series:
\begin{equation}
 \bar{\rho}(\textbf{r})=\frac{1}{V_{s}}\sum\limits_{\textbf{k}}\rho_{\textbf{k}}e^{i\textbf{k}\textbf{r}},
 \quad
 \varphi(\textbf{r})=\frac{1}{V_{s}}\sum\limits_{\textbf{k}}\varphi_{\textbf{k}}e^{i\textbf{k}\textbf{r}}.
       \label{11-5} \end{equation}
Relation (\ref{11-4}) yields $\rho_{\textbf{k}=0}=0$ and
\begin{eqnarray}
\rho_{\textbf{k} \neq 0}
&=&\int\limits_{V_{s}}\bar{\rho}(\textbf{r})e^{-i\textbf{k}\textbf{r}}d\textbf{r}=\nonumber
\\ &=& q2i\sin{(k_{z}H/2)}
\sum\limits_{j=1}^{N_{q}}e^{-ik_{x}x_{j}-ik_{y}y_{j}}.
       \label{11-6} \end{eqnarray}

\begin{figure}
\vskip1mm
\includegraphics[width=\column]{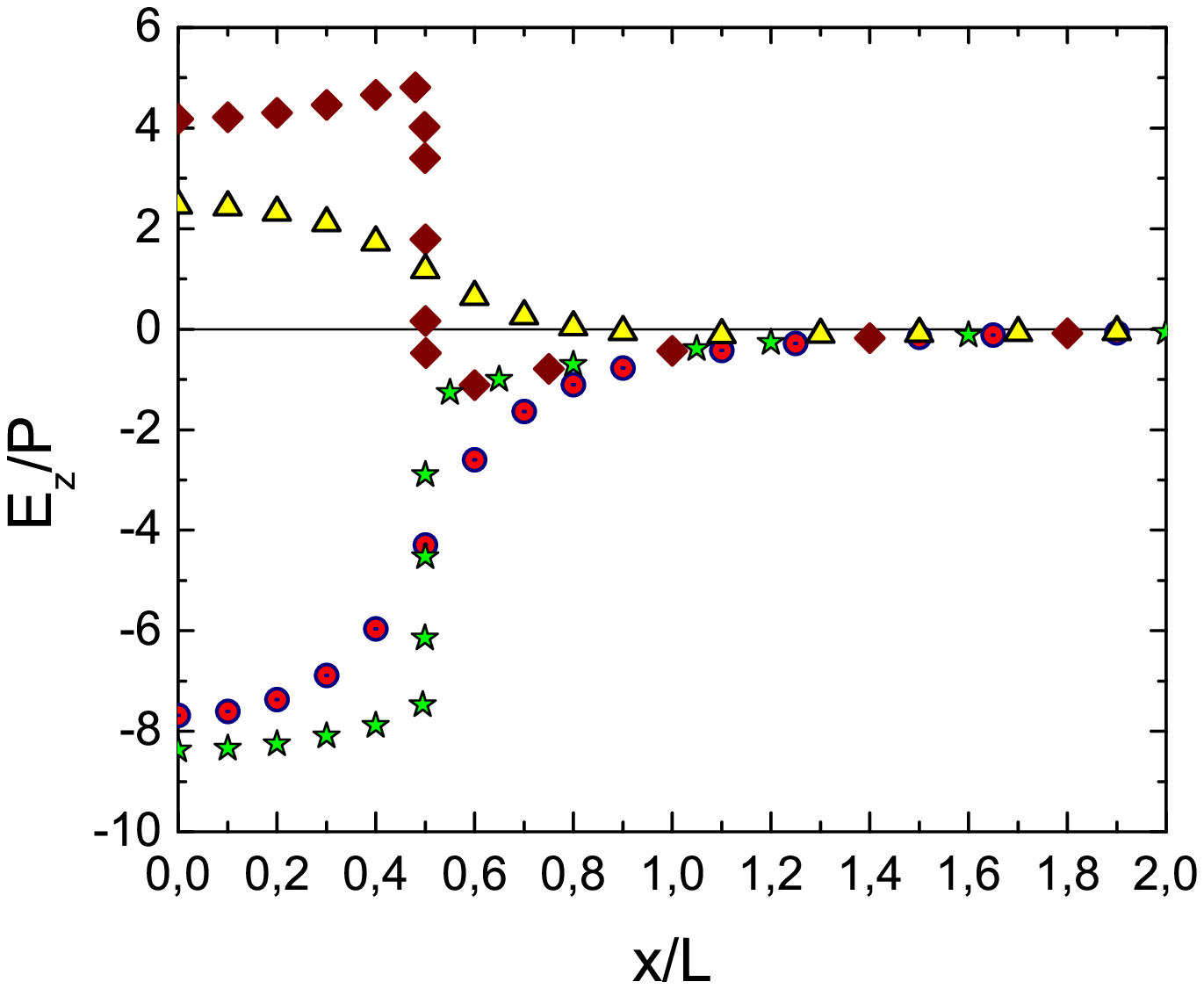}
\vskip-3mm\caption{[Color online] $E_{z}(x)$ for a dielectric plate
with $L=2H$ at $y=0$ and 1) $z=H/4$ (circles),  2) $z=0.499H$
(stars), 3) $z=0.501H$ (rhombs), 4) $z=H$ (triangles). The symmetry
implies that the function $E_{z}(y)$ at $x=0$ and $z=H/4; 0.499H;
0.501H; H$ is described by the same curves. The upper surface of the
plate corresponds to $z=0.5H$. That is, the stars and rhombs show
the field at the very boundary (under and above the surface,
respectively). The jump of $E_{z}$ on the upper surface is equal to
$4\pi P$ in agreement with BCs (\ref{bc}). }
\end{figure}

Passing to the integration and taking into account that $\frac
{q}{\triangle x \triangle y}=-P$, we get
\begin{eqnarray}
&&\rho_{\textbf{k}\neq 0}=-P2i\sin{\left ( \frac{k_{z}H}{2}\right
)}\int\limits_{-\frac{L}{2}}^{\frac{L}{2}}dx\int\limits_{-\frac{L}{2}}^{\frac{L}{2}}dy
e^{-ik_{x}x_{j}-ik_{y}y_{j}}\nonumber
\\ &&=-\frac{P8i}{k_{x}k_{y}}\sin{(k_{z}H/2)}\sin{(k_{x}L/2)}\sin{(k_{y}L/2)}.
       \label{11-7} \end{eqnarray}
At $k_{x}=0$ we set $\frac{\sin{(k_{x}L/2)}}{k_{x}}=L/2$
(analogously for $k_{y}=0$). Relation (\ref{11-3}) yields
$\varphi_{\textbf{k}\neq 0}=4\pi\rho_{\textbf{k}}/k^{2}$. Thus, we
have a solution for the potential:
\begin{eqnarray}
&&\varphi(\textbf{r})=\varphi_{0}-\frac{32\pi
P}{\Lambda^{3}}\sum\limits_{k_{x}}\frac{\sin{(k_{x}L/2)}}{k_{x}}e^{ik_{x}x}\cdot\nonumber
\\ &&\cdot \sum\limits_{k_{y}}\frac{\sin{(k_{y}L/2)}}{k_{y}}e^{ik_{y}y}
\sum\limits_{k_{z}}\frac{i\sin{(k_{z}H/2)}}{k_{x}^{2}+k_{y}^{2}+k_{z}^{2}}e^{ik_{z}z},
       \label{11-8} \end{eqnarray}
where  $k_{x}=2\pi j_{x}/\Lambda$, $k_{y}=2\pi j_{y}/\Lambda$,
$k_{z}=2\pi j_{z}/\Lambda$,  $j_{x},j_{y},j_{z}=0, \pm 1, \pm 2,
\ldots$, and $k_{x}^{2}+k_{y}^{2}+k_{z}^{2}\neq 0$. At infinity
($z=\pm \Lambda/2$), we set $\varphi=0$. Since $\varphi(z=\pm
\Lambda/2) =\varphi_{0}$, we obtain $\varphi_{0}=0$. At
$k_{\rho}\Lambda \gg 1$ the sum over $k_{z}$ in (\ref{11-8}) can be
found analytically:
\begin{eqnarray}
&&I_{z}|_{k_{\rho}\neq
0}=\sum\limits_{k_{z}}\frac{i\sin{(k_{z}H/2)}}{k_{x}^{2}+k_{y}^{2}+k_{z}^{2}}e^{ik_{z}z}=\nonumber
\\ &&=\frac{1}{2}
\sum\limits_{k_{z}}\frac{\cos{[k_{z}(H/2+z)]}-\cos{[k_{z}(H/2-z)]}}{k_{\rho}^{2}+k_{z}^{2}}=\nonumber
\\ &&= \frac{\Lambda}{4k_{\rho}}\left
(e^{-k_{\rho}|z+H/2|}-e^{-k_{\rho}|z-H/2|}\right ),
       \label{11-9} \end{eqnarray}
where $k_{\rho}=\sqrt{k_{x}^{2}+k_{y}^{2}}$. Therefore, one needs to
present $\varphi(\textbf{r})$ (\ref{11-8}) in the form $\varphi
(k_{\rho} = 0, k_{z}\neq 0)+\varphi(k_{\rho}\neq 0)$, where
$\varphi(k_{\rho}\neq 0)$ should be calculated with the help of
(\ref{11-9}). The numerical analysis shows that solutions
(\ref{11-2}) and (\ref{11-8}) coincide with a good accuracy at
$\Lambda\gsim 100L$.

For an infinite plate, we set $L=\Lambda\rightarrow \infty$ in
(\ref{11-8}). Then only the terms with $k_{\rho}=0$ are nonzero in
(\ref{11-8}). Let $k_{\rho}=10^{2}/\Lambda$. Then formula
(\ref{11-9}) is the exact one, and $k_{\rho}\rightarrow 0$. We
obtain
\begin{eqnarray}
\varphi(\textbf{r})=-\frac{8\pi
P}{\Lambda}I_{z}|_{k_{\rho}\rightarrow 0}=2\pi P\left (\left
|z+\frac{H}{2}\right |-\left |z-\frac{H}{2}\right |\right ),
       \nonumber \end{eqnarray}
which leads to (\ref{11-1}). We obtain the same result, by directly
calculating sum (\ref{11-8}). This agrees with the solution in
\cite{binder1979}. Solution (\ref{11-8}) transits smoothly into
(\ref{11-1}), as $L/H$ increases: for $L/H\gsim 100,$ the solution
is close to that in (\ref{11-1});  for $L/H=\infty,$ it coincides
with (\ref{11-1}). That is, solution (\ref{11-1}) is valid only in
the case of infinite plate.

We note that the solution for a uniformly pola\-rized cylinder of
finite length was found in \cite{moroz2007}, and the solutions for
uniformly polarized nanoparticles of different shapes were presented
in \cite{moroz2016}.

In the examples given below we consider that only $\textbf{P}_{s}$
is known and take  $\textbf{P}_{i}$ into account separately.

\subsection{Spontaneously polarized dielectric ball in vacuum}

Let the ball of radius $R$ with the  dielectric permittivity
$\varepsilon$ is uniformly spontaneously polarized:
$\textbf{P}_{s}=P_{s}\textbf{i}_{z}$. The spontaneous dipoles create
the electric field, which produces the induced polarization
\begin{equation}
\textbf{P}_{i}=\frac{(\varepsilon-1)\textbf{E}^{(int)}}{4\pi}.
       \label{2-0} \end{equation}
Consider the system as a set of spontaneous and induced dipoles in
vacuum. The field $\textbf{E}$ is determined by Eq. (\ref{11}). The
solution for a domain inside the ball is \cite{tamm}
\begin{eqnarray}
\textbf{E}^{(int)}=-\frac{4\pi \textbf{P}}{3}=-\frac{4\pi
(\textbf{P}_{i}+\textbf{P}_{s})}{3}
       \label{2-5} \end{eqnarray}
(see also Eq. (\ref{3-6e}) below, which is the solution of Eq.
(\ref{14})). Relations (\ref{2-0}) and (\ref{2-5}) yield
\begin{eqnarray}
\textbf{P}_{i}=-\frac{\varepsilon-1}{\varepsilon+2}\textbf{P}_{s},
\quad \textbf{E}^{(int)}(\textbf{r})=-\frac{4\pi}{\varepsilon
+2}\textbf{P}_{s},
       \label{2-6} \end{eqnarray}
\begin{eqnarray}
\varphi^{(int)}(\textbf{r})=\frac{4\pi}{\varepsilon
+2}\textbf{P}_{s}\textbf{r}.
       \label{2-7} \end{eqnarray}
Setting $4\pi\textbf{P}_{s}=\zeta\textbf{E}^{(int)},$ we have
$\zeta=-\varepsilon-2$, $\tilde{\varepsilon}=\varepsilon+\zeta=-2$.

The field $\varphi^{(ext)}(\textbf{r})$ outside the ball is created
by spontaneous and induced dipoles located inside the ball. This
field can be determined in the following way. The uniformly
polarized ball centered at the point $\textbf{r}=0$ can be
considered as two uniformly charged balls: the ball with charge
$-Q<0$ centered at the point $\textbf{r}_{-}=(x=y=0, z=-z_{0}/2)$
and the ball with charge $Q$ centered at the point
$\textbf{r}_{+}=(x=y=0, z=z_{0}/2)$, where $z_{0}$ is an
infinite\-simal value, and $Qz_{0}=PV=4\pi PR^{3}/3$. The ball with
charge $-Q$ creates the potential
$\varphi(\textbf{r})=\frac{-Q}{|\textbf{r}-\textbf{r}_{-}|}$ around
itself, and the ball with charge $Q$ generates the potential
$\varphi(\textbf{r})=\frac{Q}{|\textbf{r}-\textbf{r}_{+}|}$ outside
the ball. The total potential outside the polarized ball at
$z_{0}\rightarrow 0$ is
\begin{eqnarray}
\varphi^{(ext)}(\textbf{r})&=&\frac{-Q}{|\textbf{r}-\textbf{r}_{-}|}+
\frac{Q}{|\textbf{r}-\textbf{r}_{+}|}=\nonumber \\&=& \frac{4\pi
R^{3}}{3} \frac{\textbf{P}\textbf{r}}{r^{3}}=\frac{4\pi
R^{3}}{\varepsilon+2} \frac{\textbf{P}_{s}\textbf{r}}{r^{3}},
       \label{2-8} \end{eqnarray}
where
$\textbf{P}=\textbf{P}_{s}+\textbf{P}_{i}=\frac{3\textbf{P}_{s}}{\varepsilon
+2}$.  BCs (\ref{bc}) are satisfied.

We can solve the problem in a different way. If we set
$\varphi^{(int)}(\textbf{r})=A\textbf{P}_{s}\textbf{r},
\varphi^{(ext)}(\textbf{r})=B\textbf{P}_{s}\textbf{r}/r^{3}$ and
find the constants $A$ and $B$ from the BCs (\ref{bc}), we obtain
the same result.

\subsection{Spontaneously polarized dielectric placed in a spherical
conductor} Consider a grounded metallic sphere of radius $R_{m}$.
Let it contain an isotropic dielectric with uniformly distributed
over the volume spontaneous dipoles
$\textbf{d}_{s}=|q_{0}|\textbf{r}_{0}$, corresponding to the average
spontaneous polarization
$\textbf{P}_{s}=n_{s}\textbf{d}_{s}=P_{s}\textbf{i}_{z}$
($n_{s}=const$). The BC in the spherical coordinates $\rho, \theta,
\phi$ reads \cite{jackson,land8}
\begin{equation}
 \varphi(\rho= R_{m})=0.
       \label{3-1} \end{equation}
If the polarization  $\textbf{P}_{s}$ is \textit{perfectly uniform},
then, for any shape of a resonator, the problem has the known
solution
 \begin{eqnarray}
\varphi(\textbf{r})=0, \quad \textbf{E}(\textbf{r})=0,
       \label{3-14old} \end{eqnarray}
since the equation $\varepsilon \triangle \varphi=4\pi div
\textbf{P}_{s}$ with $\textbf{P}_{s}=const$ and BCs (\ref{3-1}) has
the unique solution $\varphi(\textbf{r})=0$.  However, here there is
a difficulty: On the boundary, $\textbf{P}_{s}$ decreases to zero by
jump. Therefore, $div \textbf{P}_{s}=\infty$. The solution
$\varphi(\textbf{r})=0$ neglects this property. Below, we will find
the solution within the method, which allows us to avoid this
difficulty.

It was assumed in some works that the relation
$\textbf{D}(\textbf{r})=0$ holds for a dielectric surrounded by a
metallic shell. It was not substantiated or was substantiated by
that the equality $\textbf{D}(\textbf{r})=0$ holds in a metal. In
our opinion, this reasoning is incorrect. It is well known that a
metal should be considered as a dielectric with $\varepsilon=\infty$
(see \cite{smythe} Chapt. IV, $\S 6$; \cite{land8} Chapt. II, $\S 7$
and Problem 1 after $\S 7$). In this case,
$\textbf{D}(\textbf{r})\neq 0$ inside a metal. In addition, the
condition $\textbf{D}(\textbf{r})=0$ for dielectrics leads to the
mathematical contradiction. Indeed, the equality
$\textbf{D}(\textbf{r})=0$ yields $D_{n}=0$ on the surface. If
$\varphi$ is given on a closed surface of the dielectric, then we
can find a solution of the equation $ -div
[\varepsilon(\textbf{r})\nabla\varphi(\textbf{r})]=4\pi
\bar{\rho}(\textbf{r})$ (with $\bar{\rho}(\textbf{r})$ to be known)
inside the dielectric \textit{uniquely} \cite{smythe,jackson}. This
was proved for $\varepsilon=const$ (\cite{jackson}, Chapt. 1, $\S
9$) and for $\varepsilon=\varepsilon(\textbf{r})$  (\cite{smythe},
Chapt. III, $\S 11$). If we supplement Eq. (\ref{3-1}) by the BC
$D_{n}=0$, then the problem becomes overspecified and has no
solutions for $\varphi(\textbf{r})$.

Thus, we need to solve Eq. (\ref{17}) with the BC (\ref{3-1}). The
solution for the potential is given by formula (\ref{21}). In order
to satisfy (\ref{3-1}), we need to consider ``images'' that are
reflections of dipoles in the metal. This is the most complicated
part in a problems of this kind. In our case, the simplest way to
consider the images is, apparently, the following. A real uniformly
polarized ball can be represented as two balls of radius $R$ shifted
relative each other by the size $r_{0}$ of the dipole. The first
ball is uniformly charged negatively (so that its total charge is
$Q_{-}=N_{s}q_{0}=Q_{0}$, where $N_{s}$ is the total number of
spontaneous dipoles in helium). Let its center have the coordinates
$x=y=0$, $z=-r_{0}/2$. The second ball is charged positively and has
the total charge $Q_{+}=-Q_{-}$. The coordinates of its center are
$x=y=0$, $z=r_{0}/2$. The center of the segment joining both balls
is the coordinate origin: $x=y=z=0$. We consider that these two
balls are placed in a metallic sphere so that the $(-)$ and $(+)$
balls touch the internal surface of the sphere. In this case,
$R_{m}=R+r_{0}/2$. The sphere has contact with the balls at two
opposite points, and the remaining points of the sphere are
separated from two balls by a thin layer (with a thickness of $\lsim
r_{0}/2$) of vacuum. The potential created by a uniformly charged
dielectric ball of radius $R$ at a point located at the distance
$R_{0}$ from the ball center is
\begin{eqnarray}
&&\varphi(\textbf{R}_{0})=\nonumber
\\ &&=\int\limits_{V}\frac{(\varepsilon)^{-1} g dx dy
dz}{\sqrt{(x-X_{0})^{2}+(y-Y_{0})^{2}+(z-Z_{0})^{2}}}=\nonumber \\
&&=\int\limits_{0}^{R}r^{2}dr\int\limits_{0}^{\pi}\sin{\theta}d\theta\int\limits_{0}^{2\pi}
\frac{(\varepsilon)^{-1} g d\phi
}{\sqrt{r^{2}+R_{0}^{2}-2rR_{0}\cos{\theta}}}=\nonumber\\&&=
\frac{2\pi g}{\varepsilon R_{0}}\int\limits_{0}^{R}r dr
(r+R_{0}-|r-R_{0}|).
        \label{3-2a} \end{eqnarray}
This formula yields
\begin{equation}
 \varphi(\textbf{R}_{0}) =
\left [ \begin{array}{ccc}
    \frac{2\pi g}{\varepsilon}\left (R^{2}-\frac{R_{0}^{2}}{3}\right )  & \   R_{0}\leq R,   & \\
    \frac{Q}{\varepsilon R_{0}}  & \ R_{0}\geq R. &
\label{3-2} \end{array} \right. \end{equation} Here, $g=Q/V=3Q/(4\pi
R^{3})$ is the charge density, $\varepsilon$ is the dielectric
permittivity of the ball. The dielectric weakens the field by
$\varepsilon$ times, in accordance with Eq. (\ref{17}). Solution
(\ref{3-2}) is well known \cite{tamm}.

All points at a distance of $r\leq R_{m}-r_{0}$ from the coordinate
origin belong to both balls: $(-)$-ball and $(+)$-ball. Let us
consider this domain. According to (\ref{3-2}), these two balls
create at the point $\textbf{r}$ the potential
\begin{eqnarray}
\varphi(\textbf{r})=\frac{2\pi g_{-}}{\varepsilon}\left
(R^{2}-\frac{r_{-}^{2}}{3}\right )+\frac{2\pi
g_{+}}{\varepsilon}\left (R^{2}-\frac{r_{+}^{2}}{3}\right ).
       \label{3-3} \end{eqnarray}
 Here, $g_{-}=Q_{0}/V$, $g_{+}=-g_{-}$, $V=(4\pi/3)R^{3}$, and
 $r_{-}$, $r_{+}$ are the distances from the point of observation $\textbf{r}$ to the centers of the
$(-)$ and $(+)$ balls, respectively:
\begin{equation}
r_{-}^{2}= r^{2}+(r_{0}/2)^{2}-rr_{0}\cos{(\pi-\theta)},
       \label{3-4} \end{equation}
\begin{equation}
r_{+}^{2}= r^{2}+(r_{0}/2)^{2}-rr_{0}\cos{\theta}.
       \label{3-5} \end{equation}
In this case, the vector $\textbf{r}$ is directed from the
coordinate origin to the point of observation, and
$\theta=(\widehat{\textbf{i}_{z},\textbf{r}})=(\widehat{\textbf{r}_{0},\textbf{r}})$.
Formulae (\ref{3-3})--(\ref{3-5}) yield the exact solution
\begin{eqnarray}
\varphi(\textbf{r})=-\frac{Q_{0}rr_{0}\cos{\theta}}{\varepsilon
R^{3}}=-\frac{Q_{0}\textbf{r}\textbf{r}_{0}}{\varepsilon R^{3}}
=\frac{4\pi \textbf{P}_{s}\textbf{r}}{3\varepsilon},
       \label{3-6} \end{eqnarray}
which is also well known \cite{tamm} (in \cite{tamm}, formula
(\ref{3-6}) was deduced without $\varepsilon$, since the response of
the medium to an \textit{external} field was calculated). It is
essential that relation (\ref{3-6}) follows directly from formula
(\ref{21}). It is easy to see, making use of the relations
(\ref{21}), (\ref{3-2a}), and (\ref{3-2}). Relation (\ref{3-6})
implies that, in the domain $r\leq R_{m}-r_{0},$ the electric field
strength is constant:
\begin{eqnarray}
\textbf{E}(\textbf{r})=-\nabla_{\textbf{r}}\varphi(\textbf{r})=\frac{Q_{0}\textbf{r}_{0}}{\varepsilon
R^{3}}=-\frac{4\pi \textbf{P}_{s}}{3\varepsilon}.
       \label{3-6e} \end{eqnarray}

According to (\ref{3-2}), the $(-)$ and $(+)$ balls located inside
the metallic sphere create on the internal surface of the sphere the
potential
\begin{eqnarray}
\varphi(\textbf{r})=\frac{Q_{-}}{\varepsilon
r_{-}}+\frac{Q_{+}}{\varepsilon r_{+}}.
       \label{3-7} \end{eqnarray}
It coincides with the potential, which is obtained, if the
$(-)$-ball and the $(+)$-ball are replaced by the point charges
$Q_{-}$ and $Q_{+}$ located at the centers of the $(-)$ and $(+)$
balls, respectively. In this case, the BC (\ref{3-1}) is easily
satisfied: potential (\ref{3-7}) can be exactly compensated on the
whole surface of the cavity, if we introduce two additional charges
that are the images of the point charges $Q_{-}$ and $Q_{+}$.

Let the point charge $Q_{-}$ be placed inside a conduc\-ting sphere
of radius $R_{m}$ at a distance of $r_{0}/2$ from its center. It is
known (see \cite{land8}, Chapt. I) that the image of such charge is
located at a distance of $l_{-}=2R_{m}^{2}/r_{0}$ from the sphere
center, and the charge of this image is $q_{-}=-Q_{-}2R_{m}/r_{0}.$
Moreover, the sphere center, charge, and image are located on the
same line, and the charge is placed between the sphere center and
the image. The potential created by the charge $Q_{-}$ and its image
is equal to zero on the whole surface of the sphere, which can be
directly verified.

The $(-)$ and $(+)$ balls induce polarization charges on the
internal surface of the metallic sphere. The field created by these
charges inside the sphere coincides with the field of images. With
regard for this, the total potential at a point $\textbf{r}$ inside
the sphere is the sum of potentials created at this point by the
$(-)$-ball, $(+)$-ball, and images of the point charges $Q_{-}$ and
$Q_{+}$. The solution for $r\leq R_{m}-r_{0}$ is as follows:
\begin{eqnarray}
&&\varphi(\textbf{r})=-\frac{Q_{0}rr_{0}\cos{\theta}}{\varepsilon
R^{3}}+ \frac{q_{-}}{\varepsilon r_{q_{-}}}+\frac{q_{+}}{\varepsilon
r_{q_{+}}}=\nonumber \\&&=
-\frac{Q_{0}rr_{0}\cos{\theta}}{\varepsilon R^{3}}-
\frac{2R_{m}Q_{0}}{\varepsilon
r_{0}r_{q_{-}}}+\frac{2R_{m}Q_{0}}{\varepsilon r_{0}r_{q_{+}}}.
       \label{3-8} \end{eqnarray}
Here, $r_{q_{-}}$ and $r_{q_{+}}$ are the distances from the point
of observation $\textbf{r}$ to the images of the charges $Q_{-}$ and
$Q_{+}$, respectively:
\begin{equation}
r_{q_{-}}^{2}= r^{2}+(l_{-})^{2}-2rl_{-}\cos{(\pi-\theta)},
       \label{3-9} \end{equation}
\begin{equation}
r_{q_{+}}^{2}= r^{2}+(l_{+})^{2}-2rl_{+}\cos{\theta},
       \label{3-10} \end{equation}
where $l_{-}=2R_{m}^{2}/r_{0}=L$ and $l_{+}=2R_{m}^{2}/r_{0}=L$ are
the distances from the image of the charge $Q_{-}$ and the image of
the charge $Q_{+}$ to the center of the spherical cavity. Since the
charges in (\ref{3-7}) are decreased by $ \varepsilon$ times, the
charges of the images in (\ref{3-8}) are also decreased by
$\varepsilon$ times.

At $r_{0}/R\ll 1$,  $r\leq R_{m}$ we have $r/L\ll 1$. Let us use
$r/L$ as a small parameter. Then relations (\ref{3-9}) and
(\ref{3-10}) yield
\begin{eqnarray}
\frac{1}{r_{q_{-}}}&=& \frac{1}{L}\left
[1-\frac{r\cos{\theta}}{L}+\frac{r^{2}}{L^{2}}\left
(\frac{3}{2}\cos^{2}{\theta}-\frac{1}{2} \right )\right.+\nonumber
\\&+& \left.
\frac{r^{3}}{L^{3}}\left (\frac{3}{2}\cos{\theta}-
\frac{5}{2}\cos^{3}{\theta}\right )+O\left (\frac{r^{4}}{L^{4}}
\right )\right ],
       \label{3-11} \end{eqnarray}
\begin{eqnarray}
\frac{1}{r_{q_{+}}}&=& \frac{1}{L}\left
[1+\frac{r\cos{\theta}}{L}+\frac{r^{2}}{L^{2}}\left
(\frac{3}{2}\cos^{2}{\theta}-\frac{1}{2} \right )+\right.\nonumber
\\&+& \left.
\frac{r^{3}}{L^{3}}\left
(\frac{5}{2}\cos^{3}{\theta}-\frac{3}{2}\cos{\theta} \right )+O\left
(\frac{r^{4}}{L^{4}} \right )\right ].
       \label{3-11b} \end{eqnarray}
Substituting expansions (\ref{3-11}), (\ref{3-11b}) in (\ref{3-8})
and taking the relation $L=2R_{m}^{2}/r_{0}$ into account, we get
\begin{eqnarray}
\varphi(\textbf{r})&=&\frac{Q_{0}rr_{0}\cos{\theta}}{\varepsilon}\left
(\frac{1}{R_{m}^{3}}-\frac{1}{R^{3}} \right )+\nonumber \\&
+&\frac{Q_{0}r^{3}r_{0}^{3}}{8R^{7}_{m}\varepsilon}
(5\cos^{3}{\theta}-3\cos{\theta})+O\left (\frac{r^{5}}{L^{5}} \right
).
       \label{3-12} \end{eqnarray}
In view of the formulae
$rr_{0}\cos{\theta}=\textbf{r}\textbf{r}_{0}$, $R_{m}=R+r_{0}/2$ and
the smallness of $r_{0}/R$, relation (\ref{3-12}) yields finally:
\begin{eqnarray}
\varphi(\textbf{r})=-\frac{Q_{0}\textbf{r}\textbf{r}_{0}}{\varepsilon
R^{3}}\frac{3r_{0}}{2R}+O\left (\frac{r_{0}^{3}}{R^{3}} \right ),
       \label{3-13} \end{eqnarray}
 \begin{eqnarray}
\textbf{E}(\textbf{r})=-\nabla_{\textbf{r}}\varphi(\textbf{r})\approx
\frac{Q_{0}\textbf{r}_{0}}{\varepsilon
R^{3}}\frac{3r_{0}}{2R}=-\frac{2\pi\textbf{P}_{s}}{\varepsilon
}\frac{r_{0}}{R}.
       \label{3-14} \end{eqnarray}
Two last formulae imply that the spherical conductor decreases
potential (\ref{3-6}) and the field strength (\ref{3-6e})
approximately by $R/r_{0}$ times. For example, the known mechanisms
of polarization of He II give the value of $r_{0}$ comparable with
(or much less of) the interatomic distance. Taking the realistic
values $R\sim 1\,$cm and $r_{0}\sim 10\,\mbox{\AA},$ we get
$R/r_{0}\sim 10^{7}$. In other words, the images almost completely
suppress the electric field $\textbf{E}$ inside a spontaneously
polarized dielectric ball.

We remark that, at $r_{0}=0$, the dielectric possesses a perfectly
uniform polarization. In this case, relations (\ref{3-13}) and
(\ref{3-14}) yield solution (\ref{3-14old}). However, for real
bodies $r_{0}$ is small, but nonzero (since $r_{0}$ is the size of
an elementary dipole, e.g., a molecule). Therefore, near the
boundary $\textbf{P}$ is nonuniform in a layer of finite thickness
$\sim r_{0}$.  The result given by (\ref{3-13}), (\ref{3-14}) is
apparently new.

Formula (\ref{3-14}) and the relation
$4\pi\textbf{P}_{s}=\zeta\textbf{E}$ yield $\zeta=-2\varepsilon
R/r_{0}$. The  dielectric coefficient $\zeta$ turns out to be very
large in modulus, which leads to the smallness of $\textbf{E}$. In
this case,
$\textbf{D}=(\varepsilon+\zeta)\textbf{E}=\varepsilon\textbf{E}
(1-2R/r_{0})$.

We now determine the field inside a spherical conductor in a thin
layer of width $r_{0}$ near the sphere surface. This layer is
divided into two regions: the region lying outside the $(-)$-ball
and inside the $(+)$-ball (or vice versa) and the region lying
outside the $(-)$ and $(+)$ balls.

For the region outside the $(-)$-ball and inside the $(+)$-ball, the
above formulae give
\begin{eqnarray}
&&\varphi(\textbf{r})=\frac{2\pi g_{-}}{\varepsilon}\left
(R^{2}-\frac{r_{-}^{2}}{3}\right )+\frac{Q_{+}}{\varepsilon r_{+}}+
\frac{q_{-}}{\varepsilon r_{q_{-}}}+\nonumber
\\&&+\frac{q_{+}}{\varepsilon r_{q_{+}}}\approx
\frac{3Q_{0}}{2\varepsilon
R}+\frac{Q_{0}\textbf{r}\textbf{r}_{0}}{2\varepsilon
R^{3}}\left (1-\frac{3r_{0}}{R}\right )- \nonumber \\
&&-\frac{Q_{0}\textbf{r}\textbf{r}_{0}}{2\varepsilon
r^{3}}-\frac{Q_{0}(4r^{2}+r_{0}^{2})}{8\varepsilon
R^{3}}-\frac{Q_{0}}{\varepsilon r}-\nonumber
\\&&-\frac{Q_{0}r^{2}_{0}(3\cos^{2}{\theta}-1)}{8\varepsilon r^{3}}
+O\left (\frac{r_{0}^{3}}{R^{3}} \right ),
       \label{3-15} \end{eqnarray}
 \begin{eqnarray}
\textbf{E}(\textbf{r})\approx
-\frac{3Q_{0}(\textbf{r}\textbf{r}_{0})\textbf{i}_{\textbf{r}}}{2\varepsilon
r^{4}}+\frac{Q_{0}\textbf{i}_{\textbf{r}}(r^{3}-R^{3})}{\varepsilon
R^{3}r^{2}}, \ \textbf{i}_{\textbf{r}}=\frac{\textbf{r}}{r}.
       \label{3-16} \end{eqnarray}
Since $r\approx R$ for this region, it is seen that the strength
$\textbf{E}$ is comparable by magnitude with strength (\ref{3-6e})
of the problem without a resonator.

For the region between the dielectric and the conductor (outside the
$(-)$ and $(+)$ balls), we have
\begin{eqnarray}
\varphi(\textbf{r})&=&\frac{Q_{-}}{\varepsilon
r_{-}}+\frac{Q_{+}}{\varepsilon r_{+}}+ \frac{q_{-}}{\varepsilon
r_{q_{-}}}+\frac{q_{+}}{\varepsilon r_{q_{+}}}\approx \nonumber
\\&\approx & \frac{Q_{0}\textbf{r}\textbf{r}_{0}}{\varepsilon
R_{m}^{3}}-\frac{Q_{0}\textbf{r}\textbf{r}_{0}}{\varepsilon r^{3}},
       \label{3-17} \end{eqnarray}
 \begin{eqnarray}
\textbf{E}(\textbf{r})\approx
-\frac{Q_{0}\textbf{r}_{0}}{\varepsilon
R_{m}^{3}}+\frac{Q_{0}\textbf{r}_{0}}{\varepsilon
r^{3}}-\frac{3Q_{0}(\textbf{r}\textbf{r}_{0})\textbf{i}_{\textbf{r}}}{\varepsilon
r^{4}}.
       \label{3-18} \end{eqnarray}
Here, the field strength is also comparable with (\ref{3-6e}). In
this case, strength (\ref{3-16}) is directed radially, (\ref{3-18})
has the radial and $z$ components, and strength (\ref{3-6e}) is
directed along the $z$-axis.

In the region $r\leq R_{m},$ the potential is continuous, whereas
the field strength undergoes a jump on the surface of the dielectric
and the internal surface of the conductor. It is seen from
(\ref{3-17}) that the BC (\ref{3-1}) is satisfied.

Thus, the electric field is strong (comparable with the field in the
absence of a conductor) only in the narrow space of thickness $\leq
r_{0}$ near the internal surface of the conductor. In the remaining
volume inside the conductor (i.e., in almost whole volume of the
dielectric), the field $\textbf{E}$ is almost completely suppressed.
This effect is caused by the presence of the conductor around the
dielectric. It has a simple visual explanation. Without a conductor,
the field $\textbf{E}$ inside the dielectric is uniform (see
(\ref{3-6e})). The field created by the conductor in the region
$r\leq R_{m}$ coincides with the field of images. But the images are
remote from the dielectric by the distance $L= 2R^{2}_{m}/r_{0}$,
which is much larger than the size $2R$ of the dielectric.
Therefore, the field created by the images inside the dielectric is
almost uniform, but is directed against the intrinsic field
(\ref{3-6e}) of the dielectric and compensates it. If the
compensation would be absent, then the condition $\varphi=0$ would
not be satisfied: the condition $\varphi=const$ requires
$\textbf{E}_{t}=0$ on the internal surface of the conductor. Since
the field $\textbf{E}$ is uniform inside the dielectric, the
condition $\textbf{E}_{t}=0$ on the surface requires that
$\textbf{E}\approx 0$ in the whole volume of the dielectric.

Note that a decrease in the electric energy of the dielectric
because of the location of this dielectric in a spherical conductor
is equal to the work (with opposite sign) that must be performed in
order to transport two metallic hemispheres from infinity and to
enclose a polarized dielectric by these hemispheres.

As a rule, a dielectric is in the gravity field. Therefore, such
dielectric is slightly spontaneously polarized. If the walls of a
vessel are vertical, and the bottom is horizontal, the polarization
is uniform. In this case, the electric field is weak, but measurable
\cite{mt2010}. However, if the shape of a vessel is a sphere, then
the polarization should have the $z$- and $\textbf{r}$-components
and, therefore, should be nonuniform. In other words, it is
apparently impossible to get the uniform spontaneous polarization
$\textbf{P}_{s}$ in a spherical vessel.

We mentioned above that in experiments \cite{ryb2004} the resonator
was placed in a metallic shell. The polarization of helium in those
experiments is due to the second sound and, therefore, should be
nonuniform. In this case, we  expect that the shell weakens the
field $\textbf{E}$ only by several times or by one order of
magnitude.

\vspace*{-5mm} \rezume{%
М.Д. Томченко} {ЕЛЕКТРИЧНЕ ПОЛЕ ТА ЕЛЕКТРИЧНІ СИЛИ У СПОНТАННО
ПОЛЯРИЗОВАНОМУ НЕПОЛЯРНОМУ ІЗОТРОПНОМУ ДІЕЛЕКТРИКУ} {Виходячи з
мікроскопічних рівнянь Максвелла, ми будуємо метод опису
електричного поля  в спонтанно поляризованому ізотропному
неполярному діелектрику. Ми знаходимо розв'язок для електричного
поля $\textbf{E}(\textbf{r})$ для кількох характерних прикладів.
Крім того,  ми узагальнюємо формулу Гельмгольця для електричної
сили, яка діє на елемент об'єму діелектрика, враховуючи внесок
спонтанної поляризації. }


\end{document}